\documentclass[twocolumn]{aastex631}

\renewcommand{\texttt}[1]{%
  \begingroup
  \ttfamily
  \begingroup\lccode`~=`/\lowercase{\endgroup\def~}{/\discretionary{}{}{}}%
  \begingroup\lccode`~=`[\lowercase{\endgroup\def~}{[\discretionary{}{}{}}%
  \begingroup\lccode`~=`.\lowercase{\endgroup\def~}{.\discretionary{}{}{}}%
  \catcode`/=\active\catcode`[=\active\catcode`.=\active
  \scantokens{#1\noexpand}%
  \endgroup
}
\usepackage{lipsum}
\usepackage{multirow}

\newcommand\spitzer{{\it Spitzer}}
\newcommand\acosmos{A$^3$COSMOS}

\shorttitle{ALMA sources among COSMOS/SMUVS galaxies at $z>2$}
\shortauthors{Suzuki et al.}
\graphicspath{{./}{figures/}}
\begin{document}

\title{ALMA sub-/millimeter sources among {\it Spitzer} SMUVS galaxies at $z>2$ in the COSMOS field}

\correspondingauthor{Tomoko L. Suzuki}
\email{tomoko.suzuki@ipmu.jp}

\author[0000-0002-3560-1346]{Tomoko L. Suzuki}\altaffiliation{Canon Foundation in Europe Research Fellow}
\affiliation{Kavli Institute for the Physics and Mathematics of the Universe (WPI),The University of Tokyo Institutes for Advanced Study, The University of Tokyo, Kashiwa, Chiba 277-8583, Japan}
\affiliation{Kapteyn Astronomical Institute, University of Groningen, P.O. Box 800, 9700AV Groningen, the Netherlands} 

\author[0000-0001-8289-2863]{Sophie E. van Mierlo}
\affiliation{Kapteyn Astronomical Institute, University of Groningen, P.O. Box 800, 9700AV Groningen, the Netherlands} 

\author[0000-0001-8183-1460]{Karina I. Caputi}
\affiliation{Kapteyn Astronomical Institute, University of Groningen, P.O. Box 800, 9700AV Groningen, the Netherlands}



\begin{abstract}
Sub-millimeter observations reveal the star-formation activity obscured by dust in the young Universe.
It still remains unclear how galaxies detected at sub-millimeter wavelengths are related to ultraviolet/optical-selected galaxies in terms of their observed quantities, physical properties, and evolutionary stages. 
Deep near- and mid-infrared observational data are crucial to characterize the stellar properties of galaxies detected with sub-millimeter emission.   
In this study, we make use of a galaxy catalog from the \spitzer\ Matching Survey of the UltraVISTA ultra-deep Stripes. 
By cross-matching with a sub-millimeter source catalog constructed with the archival data of the Atacama Large Millimeter/submillimeter Array (ALMA), 
we search for galaxies at $z>$ 2 with a sub-millimeter detection in our galaxy catalog. 
We find that the ALMA-detected galaxies at $z>$ 2 are systematically massive and have redder $K_s$--[4.5] colors than the non-detected galaxies.   
The redder colors are consistent with the larger dust reddening values of the ALMA-detected galaxies obtained from SED fitting. 
We also find that the ALMA-detected galaxies tend to have brighter 4.5~$\mu$m magnitudes.  
This may suggest that they tend to have smaller mass-to-light ratios, and thus, to be younger than star-forming galaxies fainter at sub-millimeter wavelengths with similar stellar masses. 
We identify starburst galaxies with high specific star-formation rates among both ALMA-detected and non-detected SMUVS sources. 
Irrespective of their brightness at sub-millimeter wavelengths, these populations have similar dust reddening values, which may suggest a variety of dust SED shapes among the starburst galaxies at $z>2$.
\end{abstract}



\section{Introduction} \label{sec:intro}

Star-formation in galaxies is accompanied by dust production, and the ultraviolet (UV) light from young and massive stars in star-formation regions is absorbed by dust and re-emitted as thermal emission at infrared (IR) wavelengths.
It is crucial to trace both the dust-obscured and unobscured components of the galaxy spectrum in order to account for the total star-formation activity in galaxies in an unbiased way. 
The fraction of dust-obscured star-formation in the cosmic star-formation rate density is said to increase with redshifts up to $z\sim2$ (e.g., \citealt{takeuchi05,burgarella13}, and \citealt{madau14} for review). 
Furthermore, recent sub-millimeter (mm) observations suggest that $\sim$ 40--50\% of the total star-forming activity is obscured by dust at higher redshifts, such as $z\sim$ 3--4 \citep[e.g.,][]{bouwens16,zavala21} or even out to $z\sim7$ \citep{algera23}, 
The dust-obscured star-formation is considered to play an important role in galaxy formation and evolution across cosmic time.

Sub-mm bright galaxies (SMGs) were first identified with sub-mm observations with single-dish telescopes, and their observed flux densities at $\sim 850\, \mu{\rm m}$ -- 1~mm are typically larger than a few mJy \citep[e.g.,][]{smail97, barger98, hughes98}.
Follow-up observations with high angular resolutions using large radio interferometries, such as the Atacama Large Millimeter/sub-millimeter Array (ALMA), allow us to pin down 
the positions of SMGs on the sky, and thus, make it easier to find their optical counterparts. 
With the multi-wavelength photometric information from optical to radio, the physical properties of SMGs are characterized \citep[e.g.,][]{hodge13,dacunha15,michalowski17,miettinen17,dudzeviciute20}. 
SMGs are said to be typically massive ($\rm log(M_*/M_\odot) > 10$) and have high star formation rates (SFRs) of a few 100--1000~${\rm M_\odot\, yr^{-1}}$ \citep[see][for recent review]{hodgedacunha20}. 
Furthermore, deep blind surveys \citep[e.g.,][]{dunlop17,aravena20,franco20,yamaguchi20} or individual observation programs targeting UV/optical-selected galaxies \citep[e.g.,][]{schinnerer16,scoville16,tadaki20} with ALMA reveal a population of galaxies with fainter sub-mm fluxes of $\lesssim1$~mJy. 
Such relatively sub-mm faint galaxies are also typically massive with $\rm log(M_*/M_\odot) > 10$ and have SFRs of $\gtrsim$ 30~${\rm M_\odot\, yr^{-1}}$.

A systematic comparison between sub-mm detected galaxies and UV/optical-selected galaxies at the same epoch is crucial to understand what galaxy populations are traced by the sub-mm observations and what is the role of such sub-mm detected galaxies on galaxy formation and evolution at high redshifts in a broader context. 
Deep near-infrared (NIR) and mid-infrared (MIR) photometric data are required to estimate the stellar properties of sub-mm detected galaxies at high redshifts accurately given their strong dust extinction \citep[e.g.,][]{dudzeviciute20,franco20,yamaguchi20}. 
This makes it possible to compare sub-mm detected galaxies and relatively sub-mm-faint galaxies systematically. 
%% SMUVS?
The \spitzer\ Matching survey of the UltraVISTA ultra-deep Stripes (SMUVS; P.I. Caputi; \citealt{ashby18,https://doi.org/10.26131/irsa401}) is a \spitzer\ \citep{werner04_spitzer} Exploration Science Program with the Infrared Array Camera (IRAC; \citealt{fazio04_irac}).  
%\footnote{the data is available at IPAC:\citet{https://doi.org/10.26131/irsa401}}
SMUVS conducted ultra-deep 3.6$\mu$m and 4.5$\mu$m imaging observations in part of the COSMOS field \citep{scoville07,https://doi.org/10.26131/irsa178}. 
The survey area is matched with that covered by the UltraVISTA ultra-deep NIR imaging observations \citep{mccracken12_ultravista}, as well as the ultra-deep Subaru imaging \citep{taniguchi07}. 
The point-source sensitivity of the IRAC 3.6~$\mu$m  and 4.5~$\mu$m channels in SMUVS reaches down to 25~mag with $4\sigma$ significance \citep{ashby18}. 
The wide-field and deep \spitzer/IRAC data by SMUVS enables us to construct a stellar mass-selected galaxy sample at $z>2$ \citep{deshmukh18}, which is expected to be insensitive to the presence of dust obscuration in galaxies. 
The SMUVS galaxy sample is suitable to systematically investigate the physical properties of dusty and non-dusty galaxies at high redshifts once the observational data at far-IR and/or sub-mm wavelengths is available.

In this study, we combined the SMUVS galaxy catalog with a public sub-mm source catalog constructed with the ALMA archival data in the COSMOS field (\acosmos; \citealt{liu19_I}) to investigate the dust-obscured star-formation activities of galaxies in the SMUVS catalog. 
At the same time, we also searched for the SMUVS sources located in ALMA maps but have no counterpart in the \acosmos\ catalog. 
By constructing two samples of galaxies both detected and undetected at sub-mm wavelengths, we aim to conduct a systematic comparison between sub-mm bright and sub-mm faint galaxies at the same epoch.

This paper is organized as follows: 
In Section~\ref{sec:a3cosmos}, we describe the galaxy catalog obtained by SMUVS briefly 
and explain the counterpart search for the SMUVS sources at $z>$ 2 in a sub-mm source catalog. 
In Section~\ref{sec:analaysis}, we explain our stacking analysis for the ALMA non-detected sources and the SED fitting analysis with the multi-wavelength data from optical to sub-mm.  
We show our results and discuss the difference between the ALMA-detected and non-detected SMUVS sources at $z>$ 2 in Section~\ref{sec:results}. 
In Section~\ref{sec:summary}, we summarize the main findings of this study. 
Throughout this paper, we use the cosmological parameters of $H_0 = 70\ {\rm km\ s^{-1}\ Mpc^{-1}}$, $\Omega_{\rm m} = 0.3$ and $\Omega_\Lambda = 0.7$. 
We assume a \citet{chabrier03} initial mass function (IMF). 
Magnitudes are given in the AB system \citep{okegunn83}.

\section{ALMA counterpart search for SMUVS galaxies}\label{sec:a3cosmos}

\subsection{SMUVS galaxy catalog}\label{sec:smuvs}

Source detection and photometry of the SMUVS sources were described in \citet{deshmukh18}. 
The source detection in SMUVS was primarily done with UltraVISTA data release 3 (DR3) $HK_s$ stack maps. 
At the positions of the detected sources in the $HK_s$ images, the photometry on the SMUVS 3.6 and 4.5~$\rm \mu m$ mosaics was obtained with a PSF-fitting technique using the DAOPHOT package on IRAF. 
When the photometry was not successfully obtained with this PSF-fitting technique, the fluxes of the IRAC channels are measured with a 2.4~arcsec diameter circular aperture at the positions of the $HK_s$ stack maps and then converted to the total fluxes by multiplying the aperture fluxes by a factor of 2.13.   
The sources detected in at least one IRAC channel are referred as the ``SMUVS sources'' \citep{deshmukh18}.
There are a total of $\sim 300,000$ SMUVS galaxies over $0.66\ {\rm deg^2}$.

%% SED fitting with Le Phare 
In this work, we use a newer version of the SMUVS catalog, which includes updated UltraVISTA photometry from DR4. 
\citet{vanmierlo22} conducted the Spectral Energy Distribution (SED) fitting for the SMUVS sources with the SED fitting code {\sc lephare} \citep{arnouts99,ilbert06}. 
They used the following photometric information available in the COSMOS field together with IRAC 3.6 and 4.5$\mu$m data from SMUVS: 
CFHT {\it u}-band; 
Subaru {\it B, V, r, $i_{p}$, $z_{p}$, $z_{pp}$, IA427, IA464, IA484, IA505, IA527, IA624, IA679, IA709, IA738, IA767, IA827, NB711}, and {\it NB816}; 
{\it HST} {\it F814W}; 
and UltraVISTA {\it Y, J, H} and $K_s$. 
The fluxes of these 26 bands were measured with a 2~arcsec diameter aperture and then converted to the total fluxes by applying point-source aperture corrections in each band \citep{deshmukh18,vanmierlo22}. 
Because the source detection and photometry in SMUVS are optimized mainly for galaxies at $z>2$, we focus on galaxies at $z>2$.
 In the following analysis, we use the photometric redshifts, stellar masses and dust reddening values $E(B-V)$ 
 of the best-fit SEDs obtained from {\sc lephare}.

\subsection{Sub-mm counterpart search with $\rm A^3 COSMOS$ catalog}\label{subsec:a3smuvs}
We use a public sub-mm source catalog from the \acosmos\ project\footnote{\url{https://sites.google.com/view/a3cosmos/home?authuser=0}} \citep{liu19_I} to search for SMUVS sources at $z>$ 2 with ALMA counterparts. 
The \acosmos\ catalog is constructed with the ALMA archival data in the COSMOS field. 
We used the sub-mm source catalog with the version of 20180801. 
There are two public catalogs of the continuum sources, namely, the blind source catalog and the prior source catalog. 
We combined the two source catalogs by matching the coordinates with a 1~arcsec searching radius as done in \citet{liu19_I}. 
Then, we conducted the cross-match with the SMUVS catalog by using the coordinates in the \acosmos\ prior source catalog for the sub-mm sources. 
In this study, we focused on the sub-mm sources detected at Band~6 ($\sim$1.2~mm) or Band~7 ($\sim870\mu{\rm m}$).  
As a result of cross-matching with a matching radius of 1~arcsec, we found 157 SMUVS sources at $z>2$ that have at least one counterpart in the \acosmos\ catalog. 
The median separation between the coordinates from the SMUVS and \acosmos\ catalog is 0.17~arcsec.  
The separation is smaller than 0.4~arcsec for 90\% of the cross-matched sources. 
We also visually checked whether the dust continuum emission is spatially associated to the stellar continuum emission with the ALMA maps and the {\it Ks}-band images from UltraVISTA DR4.  
We confirmed that the searching radius of 1~arcsec is reasonable for the counterpart search.

%% submm photometry for SMUVS sources 
In the following, we use the total flux in the \acosmos\ prior source catalog. 
When the sources were observed with the same band at least twice in different observing programs, 
we used the information with the closest separation from the SMUVS positions. 
Among 157 SMUVS sources with \acosmos\ counterpart, 22 sources were detected with both Band~6 and 7. 
Furthermore, six sources are detected with other bands, such as Band~3, 4 or 8, as well.

\subsection{SMUVS sources without \acosmos\ counterpart}\label{subsec:smuvs_nosubmm}
We searched for the SMUVS sources at $z=$ 2.0--5.5 that are covered by the ALMA maps in the \acosmos\ catalog but have no counterpart in the \acosmos\ catalog.  
These SMUVS sources can be regarded as galaxies with fainter sub-mm continuum flux as compared to the sources with \acosmos\ counterparts. 
The areal coverage of the ALMA maps in \acosmos\ is $79.5~{\rm arcmin^2}$ and $54.7~{\rm arcmin^2}$ for Band~6 and Band~7, respectively \citep{liu19_I}.
Because the outer region in an ALMA map has lower sensitivity due to the primary beam attenuation, we consider only the SMUVS sources in the inner regions where the primary beam response is greater than 0.5.
This leads to the exclusion of the sources that are not detected with dust emission due to shallow sensitivity limits.
We also removed the ALMA maps with smaller beam sizes of $b_{\rm maj}\lesssim0.6$~arcsec in order to ensure that the given upper limits based on the RMS level per beam can be compared with the total sub-mm fluxes of the ALMA-detected sources. 
When a source was observed multiple times with the same band, we used the map with the smallest RMS level.

We measured the aperture fluxes of the SMUVS sources without a counterpart in the \acosmos\ catalog for a stacking analysis (see Section~\ref{sec:stacking} for more detailed explanation). 
We used the ALMA maps selected as mentioned above and measured the fluxes at the positions of the SMUVS sources with an 1.5~arcsec radius aperture.  
We found that two sources have an aperture flux with $\rm S/N \geq 3$ and that the sub-mm continuum emission is spatially associated to the {\it Ks}-band images, which means that these sources can be regarded as detected at sub-mm wavelengths. 
One source has a counterpart in the \acosmos\ prior catalog only, and thus, was not identified in our counterpart search described in Section~\ref{subsec:a3smuvs}, which requires the detection in both the prior and blind source catalog.
The other source is not included in both the \acosmos\ prior and blind source catalog probably due to the faintness of the sub-mm flux and/or NIR flux used for the prior fit in \citet{liu19_I}.
We add the two sources to the sample of the ALMA-detected SMUVS sources and use the aperture fluxes as the total fluxes in the following analysis.

\subsection{Sample of SMUVS galaxies detected/non-detected with ALMA}\label{subsec_twosamples}

%% check spec-z catalog? 
We cross-matched the ALMA-detected and non-detected SMUVS sources with spectroscopic redshift catalogs available in the COSMOS filed \citep[e.g.,][]{lilly07, comparat15, kriek15, lefevre15, hasinger18}.
We found 11 ALMA-detected and 62 non-detected SMUVS sources with spectroscopic redshifts. 
We then evaluated the photometric redshift accuracy of the two samples with the fraction of outliers, which are defined as $\sigma = |z_{\rm spec}-z_{\rm phot}|/(1 + z_{\rm spec}) \ge 0.15$, 
and the normalized median absolute deviation (MAD), the median of $\sigma$ multiplied by 1.48 \citep{laigle16}. 
The ALMA-detected SMUVS sources have an outlier fraction of $18\%$ and $\sigma_{\rm MAD}=0.051$. 
As for the non-detected SMUVS sources, the outlier fraction is $15\%$ and $\sigma_{\rm MAD}=0.035$. 
Both the outlier fraction and $\sigma_{\rm MAD}$ are similar between the ALMA-detected and non-detected SMUVS sources, which means that the accuracy of the photometric redshifts from the SMUVS survey does not strongly depend on the sub-mm brightness.

We also cross-matched our samples with the {\it Chandra} X-ray point source catalog \citep{civano16}. 
Nine ALMA-detected and 27 non-detected SMUVS sources have X-ray counterpart, and among of them, three and nine sources, respectively, have spectroscopic redshifts too. 
It turned out that two out of the three ALMA-detected sources with X-ray counterparts and six out of the nine non-detected sources with X-ray counterparts were classified as the photometric redshift outliers. 
These sources are at $z_{\rm spec}<2$, and thus, active galactic nuclei (AGN) at lower redshift. 
This indicates that the SMUVS sources with X-ray counterparts are more likely to be AGN at $z<2$. 
Given the possibility that the SED fitting with galaxy templates would not work well for X-ray AGN even for ones with the correct photometric redshifts, we decided to remove all the X-ray sources in the following analysis. 
We removed the photo-$z$ outliers with no X-ray counterpart as well.

After removing the photo-z outliers and X-ray-detected sources,  
the number of the ALMA-detected and non-detected SMUVS sources becomes 150 and 1859, respectively.
Moreover, as explained in detail in Section~\ref{sec:magphys}, we removed five ALMA-detected SMUVS sources, which are considered to be poorly fitted with {\sc magphys}. 
The number of galaxies in each sample used in the following analysis is summarized in Table~\ref{tab:number}.
We assigned $4.2\sigma$ upper limits on the sub-mm fluxes of the SMUVS sources without ALMA detection according to the detection limit of $4.2\sigma$ for the prior fitting sub-mm source catalog in \citet{liu19_I}.

\begin{table}[tb]
    \caption{The number of the SMUVS sources at $z=$ 2.0--5.5 with and without ALMA detection. 
    Among 145 ALMA-detected SMUVS sources, 20 sources are detected with both Band~6 and 7. 
    As for the non-detected SMUVS sources, 87 sources have the flux upper limit in both Band~6 and 7.}
    \begin{center}
    \begin{tabular}{cccc}   \hline
     ALMA detection  &  Band~6  & Band~7 & Total  \\ 
      & ($\sim$ 1.2~mm) & ($\sim$ 870~$\mu {\rm m}$) & \\ \hline 
   Yes & 97 & 68 & 145 \\ 
   No & 1309 & 637 & 1859 \\ \hline 
    \end{tabular}
    \end{center}
    \label{tab:number}
\end{table}

\begin{figure}[tb]
    \centering
    \includegraphics[width=1.0\columnwidth]{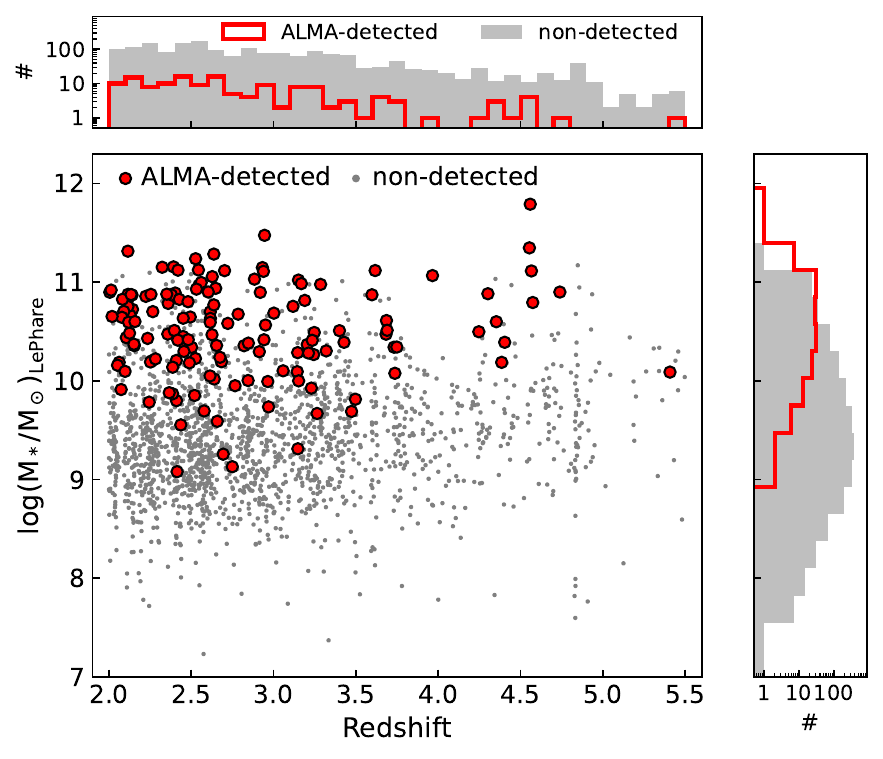}
    \caption{Stellar mass of the ALMA-detected and non-detected SMUVS sources analyzed in this study as a function of redshift. 
    Stellar masses and redshifts of both ALMA-detected and non-detected sources are from the best-fit SEDs obtained from {\sc lephare} with the 28~band photometry from {\it u}-band to 4.5~$\mu$m \citep{vanmierlo22}. 
    ALMA-detected sources are at the high-mass end of the mass distribution of the non-detected SMUVS sources whereas there is no clear difference between the redshift distributions of the two samples.  
    }
    \label{fig:massdist}
\end{figure}

\begin{figure*}
    \centering
    \includegraphics[width=1.5\columnwidth]{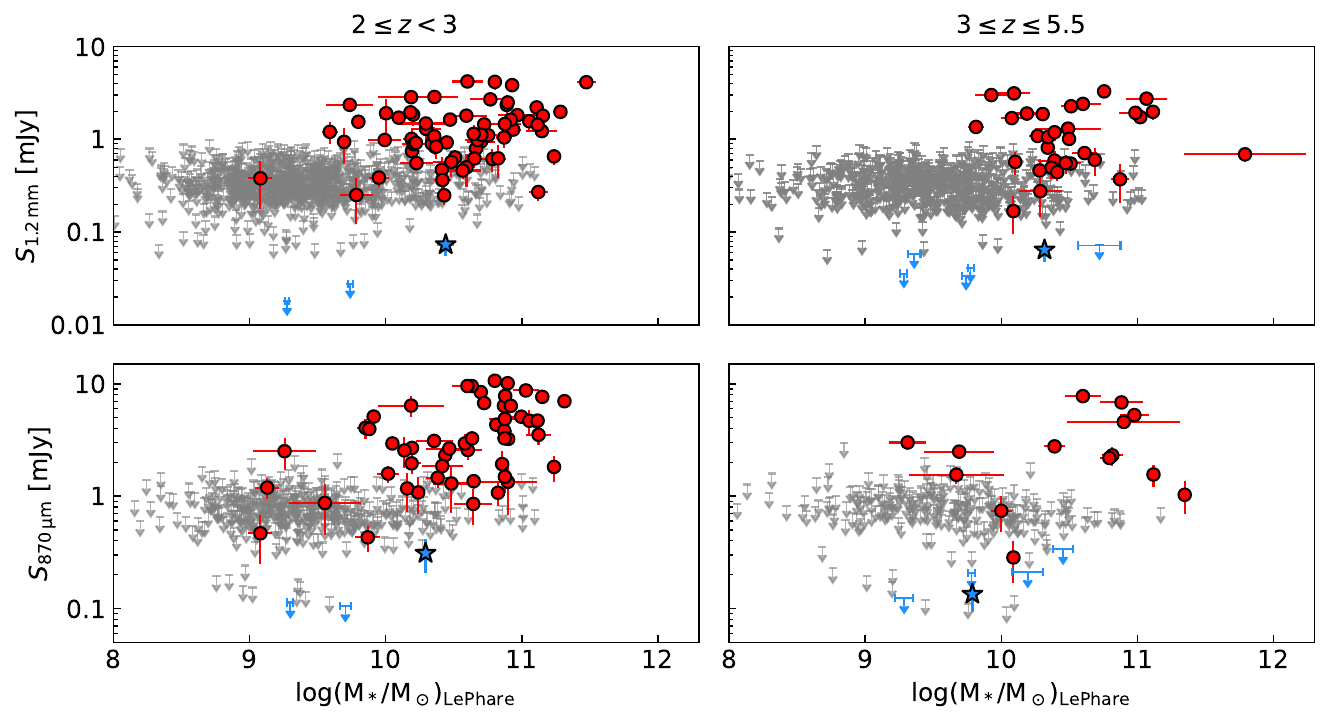}
    \caption{Continuum flux (upper limit) at $\sim$1.2~mm (top) and $\sim870\, \mu$m (bottom) as a function of stellar mass of the SMUVS sources in the two redshift bins, namely, 
    $z=$ 2.0--3.0 (left) and $z=$ 3.0--5.5 (right), analyzed in this study.
    As for the non-detected SMUVS sources, we show $4.2\sigma$ upper limits after correcting for the primary beam attenuation \citep{liu19_I}. 
    The non-detected SMUVS sources have reasonable upper limits on their continuum fluxes as compared to the individually detected SMUVS sources. 
    The results of the stacking analysis for the non-detected sources are also shown with star symbols and arrows (Section~\ref{sec:stacking}).
    }
    \label{fig:mass_flux}
\end{figure*}

Figure~\ref{fig:massdist} shows the stellar mass of the ALMA-detected and non-detected SMUVS sources at $z=$ 2.0--5.5 as a function of redshift. 
The top and right histogram shows the comparison of the redshift and stellar mass distribution of the two samples, respectively. 
We note that the stellar masses shown in Figure~\ref{fig:massdist} come from the best-fit SEDs obtained from {\sc lephare} \citep{vanmierlo22}.
Whereas the ALMA-detected and non-detected SMUVS sources have a similar redshift distribution, the ALMA-detected sources are systematically more massive ($\rm log(M_*/M_\odot) > 10$) as compared to the non-detected sources. 
The median stellar mass of the ALMA-detected and non-detected SMUVS sources is $\rm log(M_*/M_\odot)=10.51$ and $9.40$, respectively.
This trend is expected because previous sub-mm observations show that the galaxy selection based on the sub-mm brightness preferentially picks up massive star-forming galaxies \citep[e.g.,][]{dacunha15,dunlop17,yamaguchi20,dudzeviciute20}. 
In Figure~\ref{fig:mass_flux},  we show the flux (upper limit) at 1.2~mm and 870~$\mu$m of the ALMA-detected and non-detected SMUVS sources as a function of stellar mass at $z=$ 2.0--3.0 and at $z=$ 3.0--5.5, separately.  
The non-detected SMUVS sources have reasonable upper limits on their continuum fluxes as compared to the dust continuum fluxes of the individually detected sources at a given stellar mass. 
We note that the ALMA non-detected SMUVS sources include not only sub-mm faint star-forming galaxies but also galaxies with little star-formation because we do not apply any cut on the star-formation activity of galaxies.

%% Figure to show the redshift vs sub-mm flux? 

\section{Analysis}\label{sec:analaysis}

\subsection{Stacking analysis for the non-detected sources}\label{sec:stacking}

We conducted a stacking analysis for the SMUVS sources without an ALMA counterpart to investigate their average sub-mm fluxes. 
Because the pixel scales and beam sizes vary between ALMA maps from different projects, we decided to follow the stacking method with aperture fluxes applied in \citet{fudamoto20_a3cosmos}, who conducted the stacking analysis with the ALMA maps in the \acosmos\ catalog. 
As a test, we measured the aperture fluxes of the ALMA-detected SMUVS sources and confirmed that the aperture fluxes show a good agreement with the total fluxes in the \acosmos\ catalog when they are isolated. 
We conducted this test by changing the aperture radius, namely $r=$ 1.0, 1.5, and 2.0~arcsec.
The consistency with the total fluxes in the \acosmos\ catalog does not change depending on the aperture size. 
Here we use $r=$1.5~arcsec apertures.

We divided the non-detected SMUVS sources at $z=$ 2.0--5.5 with $\rm log(M_*/M_\odot) \geq 9.0$ into 18 subsamples according to their stellar masses, redshifts, and the observed wavelengths (Band~6 or 7) as summarized in Table~\ref{tab:stacksummary}.
We measured the ALMA aperture fluxes of each non-detected source at the position from the SMUVS catalog. 
We use the ALMA maps before the primary beam correction released by the \acosmos\ project\footnote{\url{https://irsa.ipac.caltech.edu/data/COSMOS/overview.html}}. 
The errors on the aperture fluxes are determined from the standard deviation of the aperture fluxes measured at 100 random positions in each ALMA map. 

Before stacking, we removed the SMUVS sources that have a close sub-mm bright source which contaminates their aperture fluxes.  
We also removed passive galaxies, which are considered to be intrinsically faint at sub-mm, in order to increase the signal-to-noise ratio (S/N) of the stacking result. 
We followed the method applied in \citet{deshmukh18} to distinguish passive galaxies from star-forming galaxies. 
\citet{deshmukh18} divided the SMUVS galaxy sample at $z>2$ into subsamples of passive galaxies, dusty star-forming galaxies, and non-dusty star-forming galaxies  based on the rest-frame $u-r$ color and $E(B-V)$. 
According to their criteria, the non-detected SMUVS sources with $(u-r)_{\rm rest} > 1.3$ and $E(B-V) < 0.2$ were classified as passive galaxies and removed from the stacking analysis. 
The fraction of such sources is 5\%.

When we stacked the aperture fluxes for each subsample, the aperture fluxes were weighted according to the RMS values after correcting for the primary beam attenuation at the position of the sources \citep{fudamoto20_a3cosmos}. 
Errors on the stacked fluxes were estimated with the jackknife resampling method \citep[e.g.,][]{efron82}. 
We generated $N$ samples with the sample size of $N-1$ from a SMUVS subsample with the size of $N$. 
The {\it i}-th source was removed from the {\it i}-th jackknife sample. 
Then, we calculated a stacked flux for each jackknife sample in the same manner as done for the SMUVS subsamples.
We use the standard deviation of the stacked fluxes of the jackknife samples as an error on the stacked flux of the SMUVS subsample. 
%%%

Table~\ref{tab:stacksummary} summarizes the results of the stacking analysis. 
Four out of the 18 subsamples show stacked continuum fluxes with ${\rm S/N}>3$. 
As expected, subsamples consisting of galaxies with higher stellar masses and at lower redshift tend to have higher S/N.
The remaining subsamples show the stacked fluxes with $<3\sigma$ such that we adopt $3\sigma$ flux upper limits for them.  
The relation between the stacked fluxes and stellar masses of 18 subsamples is shown in Figure~\ref{fig:mass_flux}. 
The stacking-detected subsamples have $\sim$ 1~dex fainter fluxes as compared to the individually detected SMUVS with similar stellar masses. 
The comparison between the stacked subsamples and individually detected sources would indicate the large scatter of sub-mm continuum fluxes of star-forming galaxies even at the same stellar mass.

\begin{deluxetable*}{c|c|cccc|cccc} 
\centerwidetable
 \caption{Summary of the stacking analysis for the non-detected SMUVS sources at $z=$ 2.0--5.5. When S/N$<3$, the $3\sigma$ upper limits are assigned. The redshifts and stellar masses of the subsamples are estimated by taking a weighted average as done for the aperture fluxes. \label{tab:stacksummary}} 
   \tablehead{
   \multicolumn{2}{c}{Bins}  & \multicolumn{4}{c}{Band-6 (1.2~mm)} & \multicolumn{4}{c}{Band-7 (870~$\mu$m)} \\
   \colhead{$z$} & \colhead{$\rm log(M_*/M_\odot)$} & \colhead{\#} & \colhead{Stacked flux [mJy]} & \colhead{$z$} & \colhead{$\rm log(M_*/M_\odot)$} & \colhead{\#} & \colhead{Stacked flux [mJy]} & \colhead{$z$} & \colhead{$\rm log(M_*/M_\odot)$}}  
   \startdata
   \multirow{3}{*}{2.0--3.0} &  $>$10.0  & 83 &  $0.073\pm0.018$  & $2.49\pm0.05$ &  $10.44\pm0.06$ & 49 & $0.31\pm0.10$ & $2.34\pm0.06$ &  $10.29\pm0.03$ \\
   &  9.5--10.0  & 178 & $<0.028$  & $2.55\pm0.04$ & $9.74\pm0.02$  &  89 & $<0.105$ & $2.53\pm0.06$ & $9.70\pm0.04$  \\
   &  9.0--9.5   &  294 & $<0.018$ & $2.48\pm0.03$ & $9.28\pm0.01$  &  148  & $<0.113$ & $2.46\pm0.04$ & $9.30\pm0.02$  \\  \hline 
   \multirow{3}{*}{3.0--4.0} &  $>$10.0   & 49 &  $0.064\pm0.016$  & $3.45\pm0.09$  &  $10.32\pm0.05$  & 30 & $<0.341$ & $3.71\pm0.06$ &  $10.45\pm0.07$ \\
   &  9.5--10.0  & 117 & $<0.041$  & $3.39\pm0.05$ & $9.77\pm0.02$  &  45 & $<0.209$ & $3.54\pm0.07$ & $9.78\pm0.03$  \\
   &  9.0--9.5   &  124 & $<0.036$ & $3.46\pm0.06$ & $9.28\pm0.03$ &  48  & $<0.125$ & $3.56\pm0.09$ & $9.29\pm0.07$  \\ \hline
   \multirow{3}{*}{4.0--5.5} &  $>$10.0   & 19 &  $<0.072$  & $4.88\pm0.11$ &  $10.72\pm0.15$ & 16 & $<0.212$ & $5.31\pm0.17$ &  $10.19\pm0.11$ \\
   &  9.5--10.0  & 47 & $<0.034$  & $4.57\pm0.05$ & $9.74\pm0.03$ &  26 & $0.134\pm0.041$ & $4.66\pm0.25$ & $9.79\pm0.03$  \\
   &  9.0--9.5   &  38 & $<0.058$ & $4.70\pm0.14$ & $9.36\pm0.05$ &    \multicolumn{4}{c}{--}    
    \enddata
\end{deluxetable*}

\subsection{SED fitting with MAGPHYS}\label{sec:magphys}
We conducted an independent SED fitting of the ALMA-detected SMUVS sources in order to take into account their stellar and dust emission properties simultaneously.
We used a SED fitting code {\sc magphys} that can fit the SEDs from the optical to radio wavelengths consistently \citep{dacunha08,dacunha15,battisti20}.

%% about magphys 
{\sc magphys} uses the stellar population synthesis models of \citet{bc03} assuming a \citet{chabrier03} IMF 
and uses the two-component dust model of \citet{charlot00} for the dust attenuation.  
The metallicity range is set to be 0.2--2.0$\times Z_\odot$,  
and the age range is 0.1--10~Gyr. 
Star-formation history is parameterized as a continuous delayed exponential function, in which the SFR rises linearly at the earlier epoch and then declines exponentially with the timescale defined by the $\gamma$ parameter ($\gamma=$ 0.075--1.5 $\rm Gyr^{-1}$).
{\sc magphys} also includes starbursts of random duration and amplitude to account for the stochastic star-formation.  
We used the {\sc magphys} high-$z$ extension version~2, which includes the 2175\AA\ feature in the dust attenuation curve \citep{battisti20}. 
The high-$z$ extension version~2 uses the intergalactic medium (IGM) absorption in the UV regime from \citet{inoue14}.

We combined the sub-mm detection(s) from ALMA with the broad-band photometry from the SMUVS catalog \citep{vanmierlo22}. 
For bands in the optical to NIR regime, we inspect the S/N in each band, 
such that if S/N $< 3$, we instead adopt a $3\sigma$ flux upper limit in that band. 
Redshifts are fixed to photometric redshifts in the SMUVS catalog.

In order to maximize the constraints on the IR SEDs, we added photometric information in the IR regime other than the ALMA data. 
We used the IR photometric catalog constructed by \citet{jin2018}.
This catalog contains multi-wavelength photometry ranging from {\it Spitzer}/IRAC 3.6~$\mu$m to the Karl G. Jansky Very Large Array (JVLA) 1.4~GHz, measured with the ``super-deblending'' technique developed by \citet{liu2018}. 
We cross-matched the coordinates of the ALMA-detected SMUVS sources with those of the sources in the super-deblended catalog of \citet{jin2018} with a searching radius of 1~arcsec.
Most of the ALMA-detected SMUVS sources ($\sim90$\%) have a counterpart in the super-deblended catalog. 
We added the photometric information from {\it Spitzer}/IRAC 5.8~$\mu$m to {\it Herschel}/SPIRE 500~$\mu$m to the photometric catalog of the ALMA-detected SMUVS sources. 
When S/N $<3$, $3\sigma$ upper limits were assigned.

%% MAGPHYS fitting results? 
In order to evaluate the goodness of the fits obtained with {\sc magphys}, we adopt the criterion introduced by \citet{battisti19}. 
They classify the sources failed to fit based on their best-fit $\chi^2$ values. 
They fit a Gaussian distribution to the lower 90~\% population of a sample and determine the mean ($\bar{\chi^2}$) and dispersion ($\sigma(\chi^2)$). 
When $\chi^2 > \bar{\chi^2} + 4\sigma(\chi^2)$, the sources are considered to be poorly fitted. 
We found that five sources in our sample have a $\chi^2$ value exceeding this criterion.
The five sources are removed in the following analysis.

\begin{figure}[tb]
    \centering
    \includegraphics[width=0.9\columnwidth]{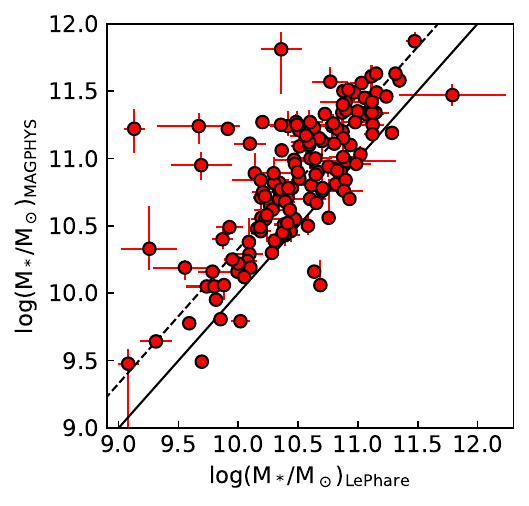}
    \caption{Comparison of the stellar masses obtained from {\sc lephare} and {\sc magphys} for the ALMA-detected SMUVS sources at $z=$ 2.0--5.5. 
    The solid line represents to the identity line.
    The dashed line represents the case when the stellar mass from {\sc magphys} is $0.25$~dex larger than that from {\sc lephare}. 
    The stellar masses from {\sc magphys} are systematically larger than those from {\sc lephare}.
    }
    \label{fig:masscomparison}
\end{figure}

Figure~\ref{fig:masscomparison} shows the comparison of the stellar masses from {\sc lephare} and {\sc magphys} of the ALMA-detected SMUVS sources at $z=$3.0--5.5. 
We find that the stellar masses from {\sc magphys} are systematically larger than those from {\sc lephare}. 
This effect was also shown in \citet{battisti19} (see also \citealt{michalowski14}). 
The difference between the two stellar mass measurements is $\sim0.25$~dex on average. 

We also conducted the SED fitting with {\sc magphys} for the four stacked subsamples with the detection of $> 3\sigma$ (Section~\ref{sec:stacking}). 
In the following analysis, we use the median IR luminosities obtained with {\sc magphys} to investigate the dust-obscured star-formation activities of the ALMA-detected SMUVS sources and the four stacking-detected subsamples. 
As for the stellar mass, we use the values obtained from {\sc lephare} for a consistency between the ALMA-detected and non-detected sources.

\subsection{Star formation rates}

The absolute UV magnitudes at rest-frame 1450~\AA\ from {\sc lephare} are available for all the SMUVS sources. 
We calculated SFRs from the rest-frame UV luminosities ($\rm SFR_{UV}$)
with the following equation from \citet{kennicutt98} scaled to a \citet{chabrier03} IMF:

\begin{equation}
    {\rm SFR_{UV}\, [M_\odot\, yr^{-1}]} = 8.8 \times 10^{-29}\, L_{\nu}\, [{\rm erg\, s^{-1}\, Hz^{-1}}], 
\end{equation}

\noindent
where $L_{\nu}$ is the luminosity at 1450~\AA. 
As for the non-detected SMUVS sources, we calculated $\rm SFR_{UV}$ after correcting for the dust extinction using $E(B-V)$ from {\sc lephare} and the \citet{calzetti00} attenuation law. 
In the following analysis, we used the dust-extinction-corrected $\rm SFR_{UV}$ as the total SFR for the non-detected SMUVS sources. 
To estimate errors on $\rm SFR_{UV}$, we used the uncertainties on the observed fluxes close to 1450~\AA\ in the rest-frame.

In the case of the ALMA-detected sources, we estimated SFRs by combining SFRs from the rest-frame UV luminosities and IR luminosities \citep[e.g.,][]{wuyts11_SFH,yamaguchi20}. 
IR luminosities are converted to $\rm SFR_{IR}$ with the \citet{kennicutt98} prescription scaled to a \citet{chabrier03} IMF and combined with $\rm SFR_{UV}$ before dust extinction correction as follows: 

\begin{eqnarray}
    {\rm SFR_{UV+IR}\, [M_\odot\, yr^{-1}]} = & & {\rm SFR_{UV, dust\, uncorr}}  \nonumber \\
    & & + 1.09 \times 10^{-10}\, L_{\rm IR} [L_\odot],
\end{eqnarray} 

\noindent
where $L_{\rm IR}$ is a median IR luminosity obtained from {\sc magphys} (Section~\ref{sec:magphys}). 
The errors on ${\rm SFR_{IR}}$ are estimated using the 16 and 84 \% percentile of the $L_{\rm IR}$ obtained from {\sc magphys}.

\section{Results and Discussion}\label{sec:results}
\subsection{NIR/MIR brightness and colors}\label{subsec:nircolor_comparison}

\begin{figure*}[htbp]
\centering\includegraphics[width=1.0\textwidth]{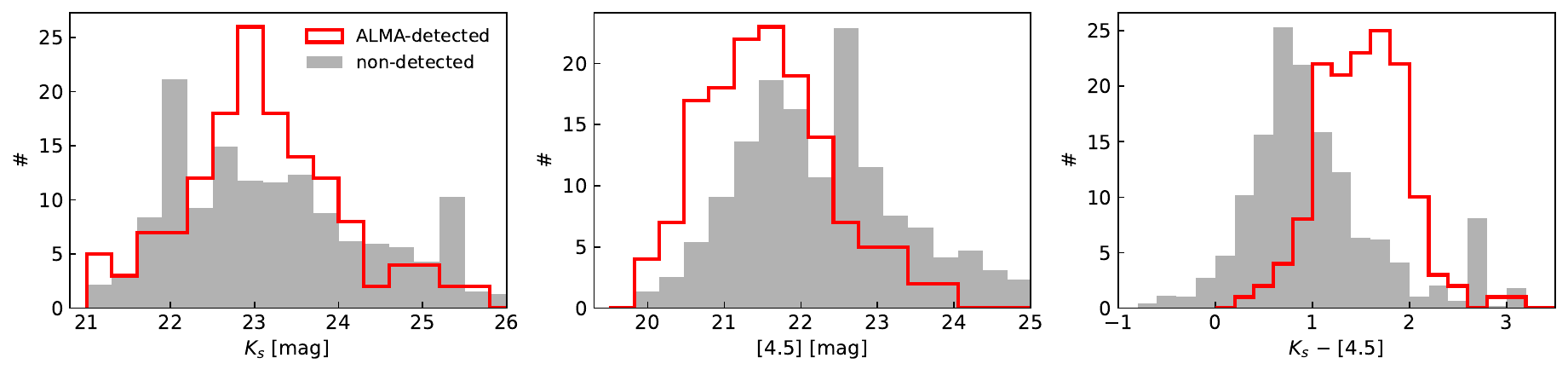}
\caption{Comparison of {\it Ks}-band magnitude, 4.5~$\mu$m magnitude, and {\it Ks}--[4.5] color between the ALMA-detected and non-detected SMUVS sources at $z=$ 2.0--5.5. 
The histograms of the non-detected SMUVS sources are weighted according to the stellar masses so that the stellar mass distribution becomes the same between the ALMA-detected and non-detected sources. 
The ALMA-detected sources at $z\ge2$ tend to be brighter at 4.5~$\mu$m and have systematically redder {\it Ks}--[4.5] colors than the non-detected sources even when considering the difference of the stellar mass distributions.} 
\label{fig:nircolor}
\end{figure*}

Figure~\ref{fig:nircolor} shows the comparison of observed quantities, namely, {\it Ks}-band magnitude, 4.5~$\mu$m magnitude, and {\it Ks}--[4.5] color, between the ALMA-detected and non-detected SMUVS sources at $z=$ 2.0--5.5.
We here gave a weight to each non-detected SMUVS source according to its stellar mass so that the weighted stellar mass distribution of the non-detected sources matches with the stellar mass distribution of the ALMA-detected sources. 
By using the weighted distribution of the non-detected sources for comparison, we can minimize the effect of the stellar mass dependency of each quantity.

The ALMA-detected and non-detected SMUVS sources at $z=$ 2.0--5.5 have similar {\it Ks}-band magnitude distributions. 
On the other hand, the 4.5~$\mu$m magnitude distribution of the ALMA-detected SMUVS sources appears to be shifted toward brighter magnitudes as compared to that of the non-detected sources. 
When comparing the {\it Ks}--[4.5] color distribution between the two samples, the ALMA-detected SMUVS sources tend to have redder colors of {\it Ks}--[4.5] $\gtrsim 1$.

Colors in the NIR and {\it Spitzer}/IRAC bands are used to select (extremely) dusty galaxies at high redshift \citep[e.g.,][]{wang12,chen16,wang16}. 
\citet{wang12} selected extremely red objects based on the {\it Ks}--[4.5] colors (KIEROs, {\it Ks}--[4.5]$>$1.6) and showed that the majority of KIEROs are massive ($\rm log(M_*/M_\odot)=$ 10--12) star-forming galaxies at $z=$ 2--4. 
Of the ALMA-detected SMUVS sources, 44\% have {\it Ks}--[4.5] $>$ 1.6 and 46\% have bluer colors of {\it Ks}--[4.5] $=$ 1.0--1.6. 
A half of them are not as extremely red as KIEROs. 
These results suggest that the {\it Ks}--[4.5] color is useful to select galaxies bright at sub-mm wavelengths at $z\ge2$, and that applying a cut at {\it Ks}--[4.5] $\sim1$ would lead to an increase in completeness.

As shown in the next Section, we find that the dust reddening values, $E(B-V)$, of the ALMA-detected SMUVS sources are systematically larger than those of the non-detected sources. 
The observed redder {\it Ks}--[4.5] colors of the ALMA-detected SMUVS sources appear to be consistent with their stronger dust extinction \citep{wang12}. 
On the other hand, the trends of the {\it Ks}-band and 4.5~$\mu$m magnitude distributions shown in Figure~\ref{fig:nircolor} seem to be difficult to explain only with the different dust extinction strength between the two samples.
Brighter 4.5~$\mu$m magnitudes of the ALMA-detected SMUVS sources may suggest that galaxies with bright sub-mm emission tend to have smaller mass-to-light ratios at $\lambda_{\rm obs}=4.5$~$\mu$m, and thus, tend to be younger as compared to those fainter at sub-mm wavelengths with similar stellar masses.

\subsection{SED properties}\label{subsec:sedcomparison}

\begin{figure*}[htb]
    \centering
    \includegraphics[width=1.7\columnwidth]{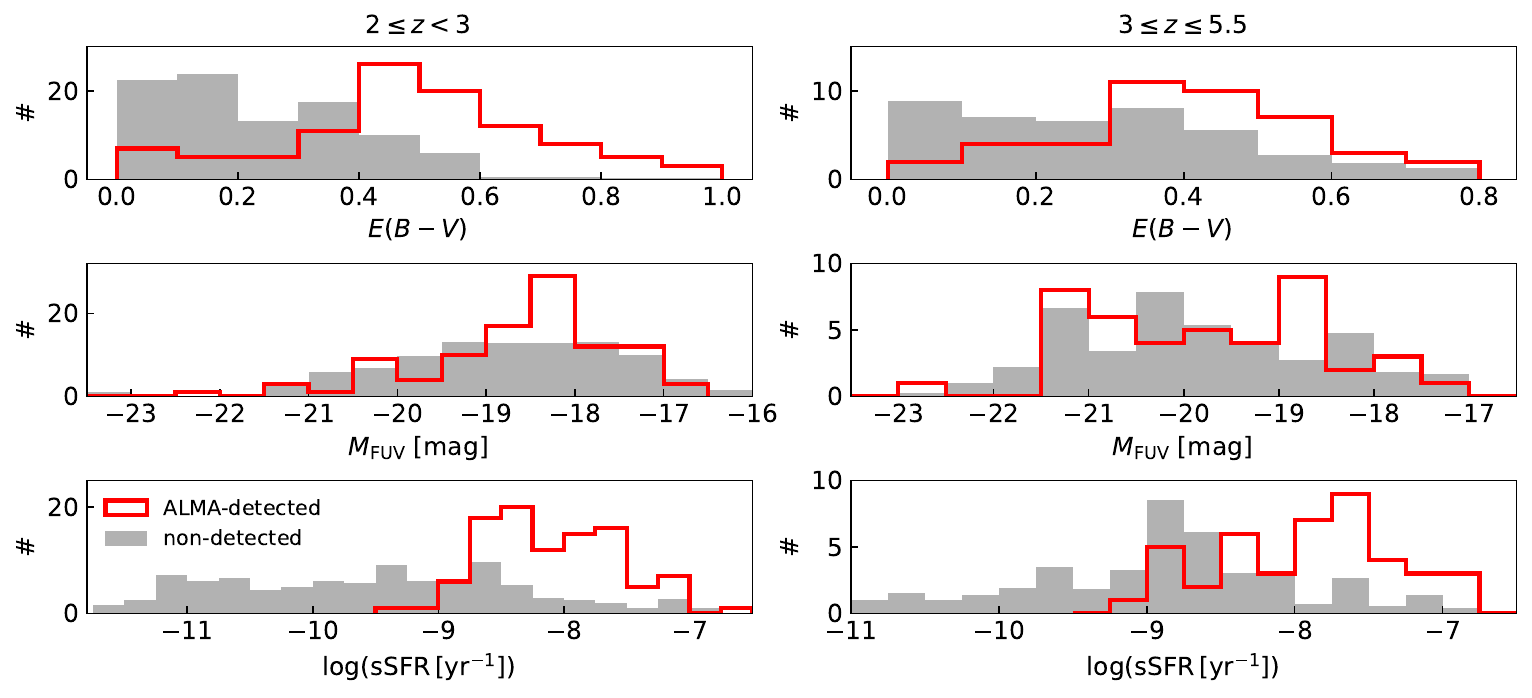}
    \caption{Comparison of the physical quantities between the ALMA-detected and non-detected SMUVS sources at $z=$ 2.0--3.0 (left) and $z=$ 3.0--5.5 (right). 
    Here the non-detected SMUVS sources are weighted according to their stellar masses as done in Figure~\ref{fig:nircolor}. 
    The ALMA-detected sources tend to be dustier and more active in star-formation than the non-detected sources. 
    }
    \label{fig:hist_sed}
\end{figure*}

In Figure~\ref{fig:hist_sed}, we compare $E(B-V)$, the dust {\it un}corrected absolute UV magnitude ($M_{\rm FUV}$), and specific SFR ($\rm = SFR/M_*$) between the ALMA-detected and non-detected SMUVS sources.
Here the histograms for the non-detected sources are weighted according to their stellar masses as done in Section~\ref{subsec:nircolor_comparison}. 
The weights are determined for each redshift bin.

The top two panels in Figure~\ref{fig:hist_sed} show that the ALMA-detected SMUVS sources tend to have larger dust reddening values than the non-detected sources.  
Most of the ALMA-detected SMUVS sources have $E(B-V) \geq 0.2$, 
and extend as far as $E(B-V)=1.0$.
As mentioned in Section~\ref{sec:stacking}, \citet{deshmukh18} used the rest-frame {\it u--r} color and $E(B-V)$ to classify SMUVS sources into three populations, namely, non-dusty star-forming galaxies ($(u-r)_{\rm rest} < 1.3$ and $E(B-V)\le0.1$), dusty star-forming galaxies ($E(B-V)\ge0.2$), and passive galaxies ($(u-r)_{\rm rest} > 1.3$ and $E(B-V)\le0.1$). 
Among the ALMA-detected SMUVS sources, only 4\% and 8\% are classified as non-dusty star-forming galaxies and passive galaxies, respectively.
This means that the classification in \citet{deshmukh18} works well for the sub-mm bright sources among the SMUVS sources and that 
{\sc lephare} appears to retrieve the dusty SEDs of the sub-mm-detected sources successfully, using only optical to IRAC photometry.

We find no clear difference between the $M_{\rm FUV}$ distributions of the ALMA-detected and non-detected sources. 
As for the sSFR distributions, sSFRs of the ALMA-detected sources appear to be biased toward higher values with $\rm log(sSFR\, [yr^{-1}]) \gtrsim -8$. 
On the other hand, the non-detected sources cover a wide range of sSFR down to $\rm log(sSFR\, [yr^{-1}]) \sim -11$.
The lack of a clear difference between the $M_{\rm FUV}$ distributions may partly reflect the fact that galaxies can be fainter in the rest-frame UV because of either stronger dust extinction or lower star-formation activity.

The ALMA-detected SMUVS sources are systematically dustier and more active in star-formation than the non-detected sources, even after taking into account the difference between the stellar mass distributions.
Such active star-formation of the ALMA-detected SMUVS sources would be consistent with their smaller mass-to-light ratios suggested in Section~\ref{subsec:nircolor_comparison}.

\subsection{SMUVS sources on $M_*$ versus SFR diagram}\label{subsec:MS}

\begin{figure*}[tb]
    \centering
    \includegraphics[width=0.8\textwidth]{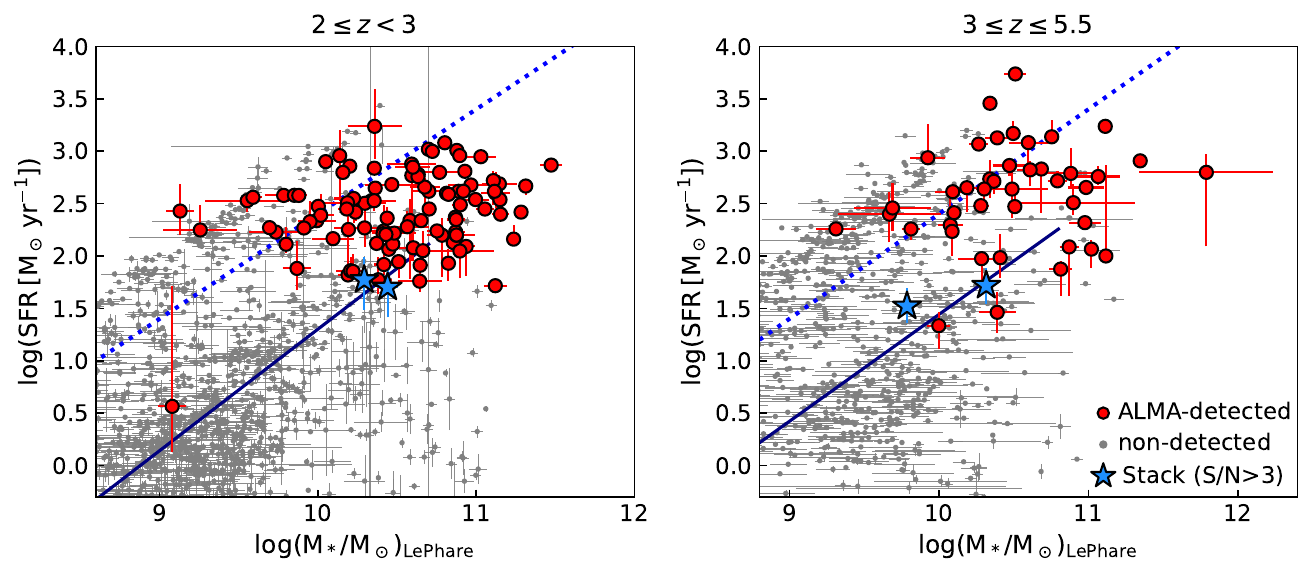}
    \caption{Stellar mass versus SFR diagram of the SMUVS sources at $z=$ 2.0--3.0 (left) and $z=$3.0--5.5 (right). 
    The stacking results with the detection greater than $3\sigma$ 
    (Section~\ref{sec:stacking}) are also shown.
    The solid line in each panel represents the star-forming main sequence at $z=$ 2--3 (left) and at $z=$ 3--4 (right) from \citet{santini17}.
    The dotted line shows the lower envelope of starburst galaxies defined by \citet{caputi21}. 
    The ALMA-detected SMUVS sources are located on and above the star-forming main sequence at the epoch. 
    }
    \label{fig:MS}
\end{figure*}

Figure~\ref{fig:MS} shows the $\rm M_*$--SFR diagram for the ALMA-detected and non-detected SMUVS sources at $z=$ 2.0--3.0 and  $z=$ 3.0--5.5.  
The non-detected SMUVS sources appear to show a bimodal distribution on this diagram.  
One sequence corresponds to the main sequence and the other corresponds to the starburst cloud located above the main sequence \citep[e.g.,][]{rodighiero11}. 
Such a bimodal distribution of SMUVS sources on the $\rm M_*$--SFR diagram was reported by \citet{caputi17} using H$\alpha$ excess galaxies at $3.9\lesssim z \lesssim 4.9$ selected from the SMUVS catalog based on the photometric excess in IRAC 3.6~$\mu$m, and later confirmed by \citet{rinaldi21} to extend at all redshifts, $z\sim$ 3.0--6.5, with an independent analysis.  
The definition of starburst galaxies is set to be $\rm log(sSFR\, [yr^{-1}]) \ge -7.6$ in \citet{caputi17,caputi21}.

The ALMA-detected SMUVS sources appear distributed across the two sequences rather than distributed on either the star-forming main sequence or the starburst cloud. 
They are located at the high-mass end of the distribution of the non-detected SMUVS sources as shown in Figure~\ref{fig:masscomparison}. 
The fraction of the ALMA-detected sources classified as starburst at $\rm log(M_*/M_\odot) \geq 9.5$ is 14\% at $z=$ 2.0--3.0 and 29\% at $z=$ 3.0--5.5. 
As for the non-detected SMUVS sources, the starburst fraction at $\rm log(M_*/M_\odot) \geq 9.5$ is 12\% at $z=$ 2.0--3.0 and 22\% at $z=$ 3.0--5.5.
When we combine the two samples in each redshift bin, the starburst fraction becomes 12\% at $z=$ 2.0--3.0 and 23\% at $z=$ 3.0--5.5.
The starburst fraction of our sample at $z=$ 3.0--5.5 is consistent with the value of 22\% obtained for galaxies with $\log(M_*/M_\odot)\ge9.5$ at $z=$ 3.0--5.0 in \citet{rinaldi21}.

In Figure~\ref{fig:MS}, we also show the stacking results with the detection greater than $3\sigma$ (Table~\ref{tab:stacksummary}). 
Given the locus of the stacking-detected subsamples on this diagram, these stacking results seem to reflect the physical properties of typical star-forming galaxies at $z=$ 2.0--5.5 at the corresponding stellar mass range.

We calculate the fraction of the dust-obscured star-formation ($f_{\rm obscured} = \rm SFR_{IR}/SFR_{UV+IR}$) for these stacking-detected subsamples as well as the individually detected SMUVS sources. 
Whereas most of the ALMA-detected SMUVS sources have $f_{\rm obscured} \sim 0.99$ irrespective of their stellar masses, the stacking-detected subsamples have $f_{\rm obscured}\sim$ 0.77--0.93. 
At a given stellar mass, the non-detected SMUVS sources appear to have a smaller contribution from the dust-obscured star-formation as compared to the individual detected sources on average.
A similar trend is reported by \citet{koprowski20} using Lyman Break Galaxies at $3\le z \le 5$ with and without ALMA detection.

\subsection{Starburst galaxies among SMUVS sources}\label{subsec:SB}

\begin{figure}[tbp]
\centering\includegraphics[width=1.0\columnwidth]{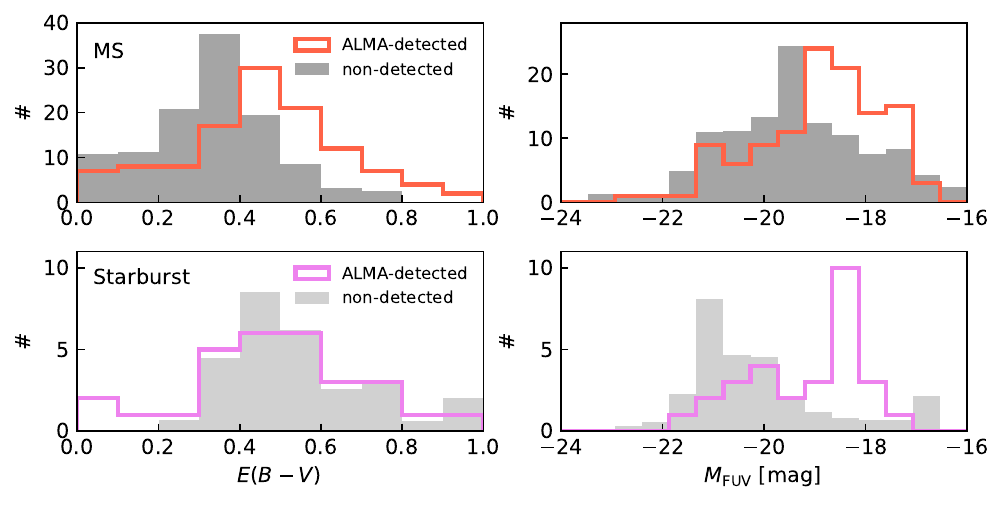}
\caption{Comparison of $E(B-V)$ (left) and $M_{\rm FUV}$ (right) between the ALMA-detected and non-detected galaxies after dividing the samples into two groups, namely, main sequence galaxies (top) and starburst galaxies (bottom).
The non-detected sources are weighted according to their stellar masses. 
Passive galaxies among the non-detected SMUVS sources are excluded based on the criteria in \citet{deshmukh18}.}
\label{fig:hist_MSSB}
\end{figure}

Figure~\ref{fig:hist_MSSB} shows the comparison of $E(B-V)$ and $M_{\rm FUV}$ between the ALMA-detected and non-detected SMUVS sources at $z=$ 2.0--5.5 after dividing the whole sample into two groups, namely, main sequence galaxies and starburst galaxies (Section~\ref{subsec:MS}). 
Here we exclude the passive galaxies classified with the \citet{deshmukh18} method (Section~\ref{sec:stacking}). 
As done in Figure~\ref{fig:nircolor} and \ref{fig:hist_sed}, 
the non-detected SMUVS sources are weighted according to their stellar masses. 
The weights are determined for the main sequence galaxies and starbursts, separately.

As for the main sequence galaxies, 
the trend seen in the $E(B-V)$ distributions is similar as what we showed for the whole sample in Figure~\ref{fig:hist_sed}. 
The difference of the $M_{\rm FUV}$ distributions between the ALMA-detected and non-detected main sequence galaxies becomes clearer than the case of the whole sample. 
The top two panels of Figure~\ref{fig:hist_MSSB} indicate that the ALMA-detected main sequence galaxies are fainter in the rest-frame UV due to their stronger dust extinction. 
They would be more dust-rich than the non-detected main sequence galaxies with similar stellar masses. 
Different sub-mm brightness between the ALMA-detected and non-detected main sequence galaxies may reflect a variety of dust masses among main sequence galaxies at a given stellar mass.

As for the starbursts, 
we find that the ALMA-detected and non-detected starbursts have similar $E(B-V)$ distributions. 
Furthermore, the non-detected starbursts tend to have larger $E(B-V)$ values than the non-detected main sequence galaxies. 
The large $E(B-V)$ values of the non-detected starbursts seem to contradict the fact that they are faint at sub-mm wavelengths. 
These results may suggest that such non-detected starbursts have higher dust temperature, which leads to fainter sub-mm fluxes at a given IR luminosity. 
Indeed, it is suggested that active galaxies above the main sequence tend to have higher dust temperatures \citep[e.g.,][]{elbaz11,magnelli14,schreiber18}.
We may see a variety of dust SED shapes among the starbursts at $z\ge2$.

\section{Summary}\label{sec:summary}

We investigated the sub-mm properties of galaxies at $z=$ 2.0--5.5 selected with the \spitzer\ SMUVS survey in the COSMOS field \citep{ashby18,deshmukh18}. 
We crossmatched the SMUVS catalog with the public sub-mm source catalog constructed with the ALMA archival data \citep[\acosmos;][]{liu19_I}. 
We also searched for SMUVS sources that are covered by the ALMA maps but have no counterpart in the \acosmos\ catalog. 
We then conducted a stacking analysis for the SMUVS sources without ALMA counterparts to investigate their average sub-mm properties.

The ALMA-detected SMUVS sources are systematically massive with $\rm log(M_{star}/M_\odot) \geq 10.0$. 
Furthermore, we find that the ALMA-detected SMUVS sources have systematically redder {\it Ks}--[4.5] colors ({\it Ks}--[4.5] $\gtrsim1.0$) than the non-detected sources even when considering the different stellar mass distributions between the two samples.
The {\it Ks}--[4.5] color together with the stellar mass information would be useful to pick up galaxies with bright sub-mm emission at $z>\ge2$. 
We also find that the ALMA-detected SMUVS sources tend to have brighter 4.5~$\mu$m magnitudes, which may suggest that galaxies with bright sub-mm emission tend to have smaller mass-to-light ratios, and thus, to be younger than those fainter at sub-mm wavelengths with similar stellar masses.

When comparing the SED properties between the ALMA-detected and non-detected SMUVS sources, we find that the ALMA-detected SMUVS sources tend to have larger $E(B-V)$ values and higher sSFRs. 
SED fitting with {\sc lephare} on the optical-to-IRAC photometry retrieves the dusty SEDs of the sub-mm-detected sources at $z\ge2$ successfully. 
Larger dust reddening values of the ALMA-detected SMUVS sources are consistent with the observed redder {\it Ks}--[4.5] colors.

On the $\rm M_*$--SFR diagram, the SMUVS sources are distributed across two regions, namely the star-forming main sequence and the starburst cloud \citep{caputi17,rinaldi21}.
Comparing $E(B-V)$ and $M_{\rm FUV}$ between the ALMA-detected and non-detected main sequence galaxies, 
we find that the ALMA-detected main sequence galaxies have larger $E(B-V)$ values and fainter $M_{\rm FUV}$, which suggests that they are likely more dust-rich than the non-detected main sequence galaxies with similar stellar masses.  
We find a different trend for the starburst galaxies. 
The non-detected starbursts have similar $E(B-V)$ values but brighter $M_{\rm FUV}$ as compared to the ALMA-detected starbursts. 
This may suggest that the non-detected starbursts have higher dust temperatures, and thus, become fainter at sub-mm wavelengths irrespective of their high star-formation activity.

High-resolution imaging observation with {\it the James Webb Space Telescope (JWST)} will enable us to investigate the rest-frame optical/NIR structures of SMUVS sources at $z>2$. 
Obtaining their stellar morphologies and color gradients with multi-band images from {\it JWST} would lead to further investigation into what causes the difference between the ALMA-detected and non-detected  sources or the difference between the main sequence galaxies and starbursts at a given stellar mass. 
The wide-field observations with NIRCam and MIRI are now being conducted in the COSMOS field (COSMOS-Web; \citealt{casey23}). 
The NIRCam imaging data in four filters covering 0.54~$\rm {deg^2}$ becomes available once the program is completed, and this will be a useful dataset for SMUVS sources.

\begin{acknowledgements}
We would like to thank the anonymous referee for a careful reading and constructive comments that improved the clarity of this paper.
TLS would like to thank John Silverman and Chris Hayward for helpful comments and suggestions.  
This work was performed with the support of the Canon Foundation in Europe. 
KC and SvM acknowledge funding from the European Research Council through the award of the Consolidator Grant ID 681627-BUILDUP. 
Kavli IPMU is supported by the World Premier International Research Center Initiative (WPI), MEXT, Japan.

This paper makes use of the following ALMA data: 
\texttt{ADS/JAO.ALMA\#2011.0.00064.S},
\texttt{ADS/JAO.ALMA\#2011.0.00097.S},
\texttt{ADS/JAO.ALMA\#2011.0.00539.S}, 
\texttt{ADS/JAO.ALMA\#2012.1.00076.S}, 
\texttt{ADS/JAO.ALMA\#2012.1.00323.S}, 
\texttt{ADS/JAO.ALMA\#2012.1.00523.S},
\texttt{ADS/JAO.ALMA\#2012.1.00536.S},
\texttt{ADS/JAO.ALMA\#2012.1.00978.S},
\texttt{ADS/JAO.ALMA\#2013.1.00034.S},
\texttt{ADS/JAO.ALMA\#2013.1.00118.S},
\texttt{ADS/JAO.ALMA\#2013.1.00151.S},
\texttt{ADS/JAO.ALMA\#2013.1.00208.S},
\texttt{ADS/JAO.ALMA\#2013.1.00884.S},
\texttt{ADS/JAO.ALMA\#2013.1.01258.S},
\texttt{ADS/JAO.ALMA\#2013.1.01292.S},
\texttt{ADS/JAO.ALMA\#2015.A.00026.S},
\texttt{ADS/JAO.ALMA\#2015.1.00055.S},
\texttt{ADS/JAO.ALMA\#2015.1.00137.S},
\texttt{ADS/JAO.ALMA\#2015.1.00207.S},
\texttt{ADS/JAO.ALMA\#2015.1.00260.S},
\texttt{ADS/JAO.ALMA\#2015.1.00379.S},
\texttt{ADS/JAO.ALMA\#2015.1.00388.S},
\texttt{ADS/JAO.ALMA\#2015.1.00540.S},
\texttt{ADS/JAO.ALMA\#2015.1.00568.S},
\texttt{ADS/JAO.ALMA\#2015.1.00664.S},
\texttt{ADS/JAO.ALMA\#2015.1.00704.S},
\texttt{ADS/JAO.ALMA\#2015.1.00928.S},
\texttt{ADS/JAO.ALMA\#2015.1.01074.S},
\texttt{ADS/JAO.ALMA\#2015.1.01105.S},
\texttt{ADS/JAO.ALMA\#2015.1.01111.S},
\texttt{ADS/JAO.ALMA\#2015.1.01171.S},
\texttt{ADS/JAO.ALMA\#2015.1.01212.S},
\texttt{ADS/JAO.ALMA\#2015.1.01495.S},
\texttt{ADS/JAO.ALMA\#2016.1.00171.S},
\texttt{ADS/JAO.ALMA\#2016.1.00279.S},
\texttt{ADS/JAO.ALMA\#2016.1.00478.S},
\texttt{ADS/JAO.ALMA\#2016.1.00804.S},
\texttt{ADS/JAO.ALMA\#2016.1.01040.S},
\texttt{ADS/JAO.ALMA\#2016.1.01208.S}.

ALMA is a partnership of ESO (representing its member states), NSF (USA) and NINS (Japan), together with NRC (Canada), MOST and ASIAA (Taiwan), and KASI (Republic of Korea), in cooperation with the Republic of Chile. 
The Joint ALMA Observatory is operated by ESO, AUI/NRAO and NAOJ.
 
\end{acknowledgements}

\software{{\sc magphys} \citep{dacunha08}, {\sc lephare} \citep{arnouts99,ilbert06}, {\sc topcat} \citep{topcat}, Astropy \citep{astropy:2013,astropy:2018},  DAOPHOT \citep{stetson87}, IRAF \citep{tody86,tody93}}


%\bibliography{reference}{}
%\bibliographystyle{aasjournal}



\end{document}